\documentclass[aps,prb,twocolumn,superscriptaddress,groupedaddress,longbibliography]{revtex4-1}
\usepackage{epsfig,amsopn}
\usepackage{graphicx}
\usepackage{color}
\usepackage{amsmath,amssymb}
\usepackage{enumerate}
\newcommand\bea{\begin{eqnarray}}
\newcommand\eea{\end{eqnarray}}
\newcommand\beq{\begin{equation}}
\newcommand\eeq{\end{equation}}

\def\nn{\nonumber}
\def\f{\frac}

\def\si{\sigma}

\def\De{\Delta}
\def\dg{\dagger}

\begin{document}
\title{Transconductance as a probe of nonlocality of Majorana fermions} 
\author{ Abhiram Soori~~}
\email{abhirams@uohyd.ac.in}
\affiliation{ School of Physics, University of Hyderabad, C. R. Rao Road, Gachibowli, Hyderabad-500046, India.}

\begin{abstract}  
Each end of a Kitaev chain in topological phase hosts a Majorana fermion. Zero bias conductance peak is an evidence 
of Majorana fermion when the two Majorana fermions are decoupled. These two Majorana fermions are separated in space and this
nonlocal aspect can be probed when the two are coupled. Crossed Andreev reflection is the evidence
of the nonlocality of Majorana fermions. Nonlocality of Majorana fermions has been proposed to be probed by 
noise measurements since simple conductance measurements cannot probe it due to the almost cancellation of currents from 
electron tunneling and crossed Andreev reflection. Kitaev ladders on the other hand host subgap Andreev states which 
can be used to control the relative currents due to crossed Andreev reflection and electron tunneling.
We propose to employ Kitaev ladder in series with Kitaev chain and show that the transconductance in this setup can be used as a
probe of nonlocality of Majorana fermions by enhancing crossed Andreev reflection over electron tunneling. 
\end{abstract}
\maketitle
\section{Introduction}
Majorana fermions~(MFs) in condensed matter systems have been of immense interest in the last couple of decades 
due to the possibility 
of realization of topological quantum computation~\cite{kitaev2001unpaired,nayak08}. MFs were predicted to exist
at the ends of semiconductor quantum wires with spin-orbit coupling proximitized with singlet superconductor in 
presence of a Zeeman field~\cite{lutchyn2010majorana,oreg2010helical}. Such quantum wires are called topological
quantum wires. Over subsequent years, many experiments have convincingly observed zero bias conductance peak in 
normal metal leads connected to topological quantum wires as predicted by the theory confirming their
realization~\cite{mourik2012,Das2012,albrecht2016,Zhang2018,aguado17}. The reason for zero bias conductance peak 
(with  a theoretically predicted conductance of $2e^2/h$) is perfect Andreev reflection mediated 
by the single MF present at zero energy. There is one MF at each end of the topological quantum wire and the two MFs can 
be coupled to form a nonlocal Dirac fermion. 
In a setup consisting of two normal metal leads attached to a superconductor, an electron incident on the superconductor
 from one normal metal can do one of the four things: reflect back, reflect back as a hole, transmit through and exit at
 the other normal metal  either as an electron or as a hole. These processes are called electron reflection~(ER), Andreev 
 reflection~(AR), electron tunneling~(ET) and crossed Andreev reflection~(CAR) respectively. 
 ET and CAR are nonlocal processes that are mediated by the nonlocal Dirac fermion formed by coupling between the 
 two MFs. ET can happen even if the superconductor in between the two normal metals is replaced by a normal metal, but 
 for CAR to happen the superconductivity is necessary. In AR, electron from one normal metal is absorbed resulting in hole 
 at the same normal metal accompanied by a Cooper pair current flowing into the topological quantum wire and this 
 process is mediated by the single MF present at the interface between the topological quantum wire and the normal metal. 
 On the other hand, in CAR, an electron from the first normal metal is absorbed resulting
 in a hole  at the second normal metal accompanied by a Cooper pair current flowing into the topological quantum wire and since
 the incident electron and the resultant hole flow in different normal metals this process is mediated by both the MFs. 
 Hence, CAR is a probe for nonlocality of MFs. In an experiment, the local conductance - the ratio of differential 
 current flowing in  one normal metal to the differential voltage applied between the normal metal and the topological 
 superconductor and the transconductance  - the ratio of differential current flowing in the second normal metal to the 
 differential voltage applied to the first normal metal maintaining the 
 topological quantum wire and the second normal metal grounded can be measured. The transconductance gets a positive 
 contribution from ET and a negative contribution from CAR. Therefore, a negative transconductance is a definite signature of 
 nonlocality of MFs. 
The nonlocality of MFs has been proposed to be probed by noise measurements~\cite{nilsson2008,Lu12,Liu13,Zocher13,
Prada17}. The reason for resorting to noise measurements is that the transconductance in a setup consisting of
two normal metal leads connected to a topological quantum wire  is negative for certain parameters 
but the magnitude is very
small despite nonlocal transport owing to the almost cancellation of electron and hole currents~\cite{nilsson2008,liu17}. 
Enhanced crossed Andreev reflection is a definite
signature  of nonlocality of MFs.  In this paper, we propose a way to probe the 
nonlocality of MFs by conductance measurements by changing the setup slightly. In the alternate setup, we show that value of 
transconductance can be made large in magnitude and negative at the same time for some specific choices of parameters. 
We employ a Kitaev ladder~\cite{nehra19} in 
series with the Kitaev chain to realize a setup to achieve this. 

\begin{figure}
  \includegraphics[width=9cm]{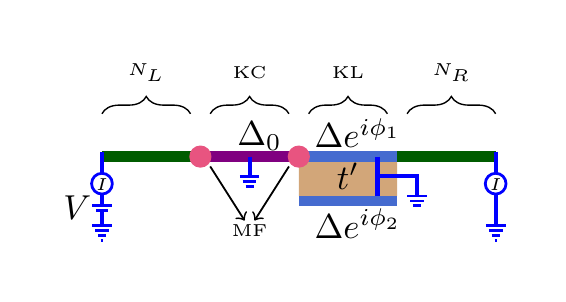}
  \caption{Schematic diagram of the setup proposed. Normal metal~($N_L$) is connected to Kitaev chain~(KC) which is then
  connected to  the Kitaev ladder~(KL) which in turn is connected to another normal metal~($N_R$). A bias voltage $V$ is applied 
  to $N_L$, while grounding KC, KL and $N_R$. Current meters denoted by I are attached through $N_L$ and $N_R$.}~\label{fig-schem}
 \end{figure}

To enhance CAR over ET is an 
 important problem and there are many methods to achieve it~\cite{deutscher2000,he14,yeyati07,soori17,nehra19}.
 One among these methods is to employ superconducting ladder which consists of two superconductors differing by a 
 superconducting phase difference forming a one-dimensional interface~\cite{soori17}. Along the interface, subgap Andreev 
 states~(SAS) exist when a substantial phase difference between the two superconductors is maintained along with a sufficiently
 large coupling. SAS are extended along the interface and they mediate ET and CAR when the interface is connected to normal
 metal leads at two ends. The relative magnitudes of ET and CAR can be changed by tuning 
 a system parameter such as chemical potential~\cite{nehra19,soori17}. We propose a setup where the Kitaev chain hosting 
 MFs at its ends is connected in series with a Kitaev ladder that hosts SAS and this entire structure is connected to normal
 metal leads as shown in Fig.~\ref{fig-schem}. The normal metal lead on the left is connected to a voltage source through a current 
 meter while the normal metal on the right is connected to the ground through a current meter. The Kitaev chain and the Kitaev ladder are 
 connected to ground. Change in currents in the two current meters as a response to change in applied voltage can be measured and hence the 
 local conductance and the transconductance can be measured.  ET and CAR in this setup are mediated by the two MFs of the Kitaev chain and
 the SAS of the Kitaev ladder. We show that in this setup, transconductance can be used as a probe of 
 nonlocality of the MFs.
 
 \section{Calculations}
 The Hamiltonian for the setup is
 \bea 
 H &=& H_{L}+H_{MF}+H_{KL}+H_{R} + H_{LM} + H_{MK} + H_{KR},\nn \\ &&~\label{eq:ham1}
 \eea
where $H_{L/R}$ is the Hamiltonian of the normal metal lead $N_{L/R}$, $H_{MF}$ is the Hamiltonian of the 
Kitaev chain hosting MFs, $H_{KL}$
is the Hamiltonian of the Kitaev ladder, $H_{LM}$ connects $N_{L}$ to Kitaev chain, $H_{MK}$ connects the Kitaev chain to 
Kitaev ladder and $H_{KR}$ connects the Kitaev ladder to $N_R$. We write the Hamiltonian using a lattice model as in 
ref.~\cite{nehra19}, though the physics can be captured even by a continuum model~\cite{soori17,thaku15}. Various terms in 
Eq.~\eqref{eq:ham1} can be written as:
\bea  
H_L &=& \sum_{n=-\infty}^{-2}[-t(c^{\dg}_{n-1}c_n+{\rm h.c.})-\mu c^{\dg}_nc_n], \nn \\
H_{MF} &=& -t_0(c^{\dg}_{-1}c_0+{\rm h.c.})-\De_0(c^{\dg}_0c^{\dg}_{-1}+{\rm h.c.}) \nn \\
&&-\mu_0 \sum_{n=1,2}(c^{\dg}_{n,\si}c_{n,\si} -\frac{1}{2}), \nn \\ 
H_{KL} &=& \sum_{\si=1,2}\sum_{n=1}^{L-1}[-t(c^{\dg}_{n+1,\si}c_{n,\si} +{\rm h.c.}) \nn \\
&&-\De(e^{i\phi_{\si}}c^{\dg}_{n+1,\si}c^{\dg}_{n,\si}  +{\rm h.c.})] \nn \\ 
&& - \mu\sum_{\si=1,2}\sum_{n=1}^L (c^{\dg}_{n,\si}c_{n,\si} -\frac{1}{2})
-t'\sum_{n=1}^L(c^{\dg}_{n,1}c_{n,2}+{\rm h.c.}), \nn \\
H_R &=& \sum_{n=L+1}^{\infty}[-t(c^{\dg}_{n+1}c_n+{\rm h.c.})-\mu c^{\dg}_nc_n], \nn \\
H_{LM}&=&-t_{LM}(c^{\dg}_{-2}c_{-1}+{\rm h.c.}), \nn \\
H_{MK}&=&-t_{MK}(c^{\dg}_0c_{1,1}+{\rm h.c.}), \nn \\
H_{KR}&=&-t_{KR}(c^{\dg}_{L+1}c_{L,1}+{\rm h.c.}). \label{eq:ham2}
\eea
Here the Kitaev chain is modeled with just two sites. In the limit $t_0=\pm\De_0$ and $\mu_0=0$, the two MFs at the two ends 
of the chain are decoupled and two-site model is a good model to capture the effects of MFs in Kitaev chain. 
Two MFs are at two ends of the Kitaev chain and are nonlocal.
To study the nonlocality of MFs, the two MFs at the two ends have to be coupled. This can be done by setting
$\De_0$ close to $t_0$, but $\De_0\neq t_0$. Then, the nonlocal fermions made from coupling of MFs are at energies 
$\pm(t_0-\De_0)$ and the quasiparticles belonging to the bulk band are at energies $\pm(t_0+\De_0)$. So, for sufficiently
large values of $t_0$ ($t_0\gg \De$) and $\De_0\sim t_0$ the quasiparticles from the bulk band of the Kitaev chain are at
energies very different from the subgap energies of the Kitaev ladder. Hence, only the MFs from the Kitaev chain participate 
in the nonlocal transport mediated by SAS of Kitaev ladder and a negative transconductance at bias energies equal to the
energies of the nonlocal fermion formed in the Kitaev chain
 is a definite signature of the nonlocality of the MFs. 

If $[\psi^e_n,~\psi^h_n]^T$ is the wavefunction at site $n$, an electron 
incident from $N_L$ with energy $E$ has a wavefunction: 
\bea 
\psi^e_n&=&e^{ik_ean}+r_ee^{-ik_ean} ~~{\rm for~~}n\le-2\nn \\
&=&t_ee^{ik_ean}~~{\rm for~~}n\ge L+1 \nn \\
\psi^h_n&=&r_he^{ik_han} ~~{\rm for~~}n\le-2 \nn \\
&=&t_he^{-ik_han}~~{\rm for~~}n\ge L+1,
\eea
where $k_ea=\cos^{-1}[-(E+\mu)/2t]$, $k_ha=\cos^{-1}[(E-\mu)/2t]$ and $a$ is the lattice spacing. The scattering coefficients 
$r_e$, $t_e$, $r_h$ and $t_h$ can be determined by writing down equation of motion using the full Hamiltonian (eq.~\eqref{eq:ham1}).
The local (nonlocal) differential conductance $G_{LL}$~($G_{RL}$) defined as the ratio of differential change in current
$dI_L$~($dI_{R}$) in lead $N_L$~($N_R$) to the differential change in applied voltage $dV_L$ at lead $N_L$ can be calculated by the 
formulas~\cite{nehra19,datta1995}:
\bea 
G_{LL}&=&\f{e^2}{h}\Big[1-|r_e|^2+|r_h|^2\f{\sin{k_ha}}{\sin{k_ea}}\Big] \nn \\
G_{RL}&=&\f{e^2}{h}\Big[|t_e|^2-|t_h|^2\f{\sin{k_ha}}{\sin{k_ea}}\Big] \label{eq:conductance}
\eea
We use the term transconductance instead of nonlocal differential conductance for $G_{RL}$.

The Kitaev ladder dispersion is 
\begin{equation}
\label{eq:disp-ladder} 
E=\nu_1\sqrt{\epsilon_k^2+{t^{\prime}}^2+\alpha_k^2+\nu_2 \cdot2t^{\prime}\sqrt{\epsilon_k^2+\alpha_k^2~\sin^2 \frac{\phi}{2}}},
\end{equation}
where $\nu_1,\nu_2=\pm 1$ represent bands formed by the hybridization of electron and hole excitations in the 
two legs of the ladder, $\phi=(\phi_1-\phi_2)$, $\epsilon_k=-(2t\cos{ka}+\mu)$ and
$\alpha_k=2\Delta\sin{ka}$. This dispersion is typically gapped and the gap closes for $\phi=\pi$ when $t'>t'_c(\mu)$, 
where $t'_c(\mu)=\Delta \sqrt{4-{\mu^2}/{(t^2-\Delta^2)}}$.

\section{Results and Analysis}
\begin{figure*}
 \includegraphics[width=8cm]{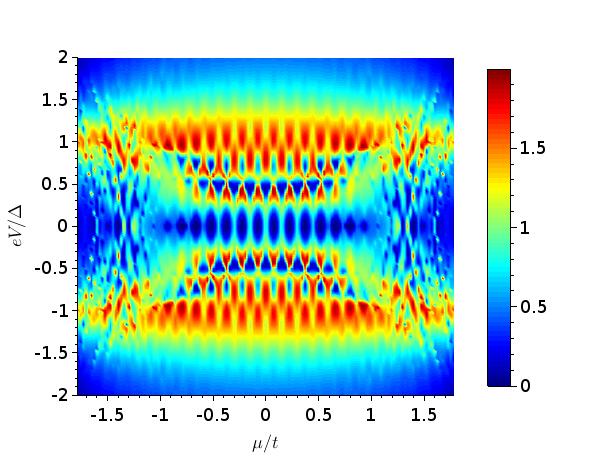}
 \includegraphics[width=8cm]{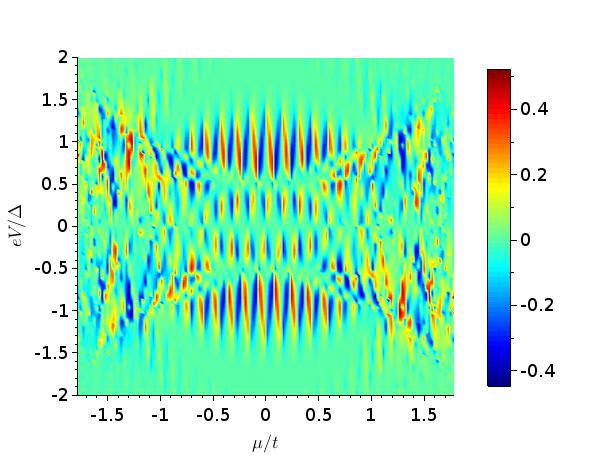}
 \caption{$G_{LL}$ (left panel) and $G_{RL}$ (right panel) in units of $e^2/h$ for the choice of parameters: $t_0=10t$, $\De_0=0.99t_0$,
$\mu_0=0$, $\De=0.1t$, $t'=3\De$, $\phi_1=0$, $\phi_2=-\pi$, $t_{LM}=0.3t$, $t_{MK}=0.3t$, $t_{KR}=t$ and $L=40$.}\label{fig09}
\end{figure*}
\begin{figure*}
 \includegraphics[width=8cm]{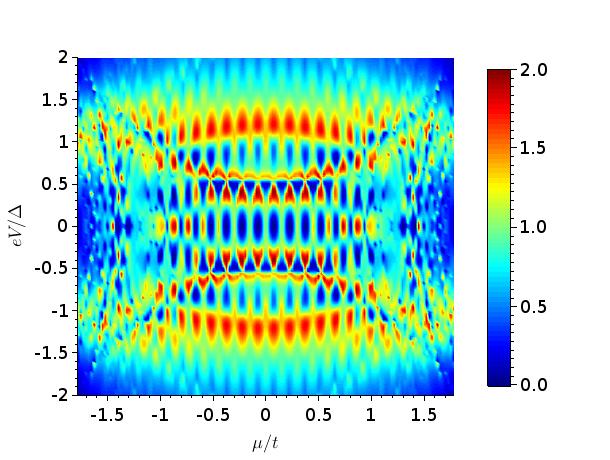}
 \includegraphics[width=8cm]{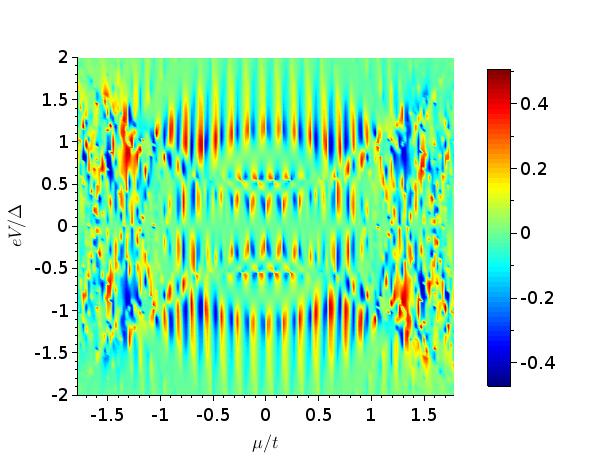}
 \caption{$G_{LL}$ (left panel) and $G_{RL}$ (right panel) in units of $e^2/h$ for the choice of parameters: $t_0=10t$, $\De_0=0.99t_0$, $\mu_0=0$,
 $\De=0.1t$, $t'=3\De$, $\phi_1=\pi$, $\phi_2=0$, $t_{LM}=0.3t$, $t_{MK}=0.3t$, $t_{KR}=t$ and $L=40$.}\label{fig09-pi}
\end{figure*}

We calculate the local conductance $G_{LL}$ and the transconductance $G_{RL}$ as functions of bias voltage $V$ and chemical potential $\mu$
for the choice of parameters: $t_0=10t$, $\De_0=0.99t_0$, $\mu_0=0$, $t_{LM}=0.3t$, $t_{MK}=0.3t$, $t_{KR}=t$ and $L=40$
in Fig.~\ref{fig09}.
Here, we choose $\De=0.1t$, $t'=3\De$, $\phi_1=0$, $\phi_2=-\pi$ so that 
there are SAS in the ladder to mediate nonlocal transport in the full energy range. The Kitaev chain is weakly coupled to 
$N_L$ ($t_{LM}=0.3t$) and to the Kitaev ladder ($t_{MK}=0.3t$). The parameters for the Kitaev chain ($\mu_0$, $t_0$ and $\De_0$) are chosen
so that only MFs participate in transport. The energy splitting between the MFs due to coupling is $2\De$ and the nonlocal fermion
states are formed at energies $\pm\De$ for this choice of parameters. We see that the local transport is dominated by AR   
and  the nonlocal transport is resonant at these energies ($eV=\pm\De$)
from Fig.~\ref{fig09}. The peaks in $G_{RL}$ have values close to $0.5e^2/h$ and the valleys have values close to $-0.5e^2/h$. The 
conductances have a periodic behavior as a function of $\mu$ due to Fabry-P\'erot interference of the SAS~\cite{nehra19,soori17,soori12}.
The periodic behavior in the transconductance $G_{RL}$ with negative values as a function of $\mu$ at $eV=\pm(t_0-\De_0)$ is a definite
signature of nonlocality of the MFs.
The Fabry-P\'erot interference condition $(k_{i+1}-k_i)aL=\pi$ determines the spacing between consecutive peaks $\mu_{i+1}-\mu_i$.
We now make two changes to the parameters of Fig.~\ref{fig09} 
by choosing $\phi_1=\pi$ and $\phi_2=0$ and plot the results in Fig.~\ref{fig09-pi}. 
The broad features due to Fabry-P\'erot interference remain the same except for a change in the details. The reason for difference in 
details is that the Kitaev chain is connected only to the upper leg of the ladder. The Kitaev chain which has zero phase forms a 
Josephson junction with the upper leg of the ladder which has a phase of $\pi$ for parameters in Fig.~\ref{fig09-pi} whereas for 
the choice of parameters in Fig.~\ref{fig09} Kitaev chain does not form a Josephson junction with the upper leg of the ladder.

\begin{figure}
 \includegraphics[width=4cm]{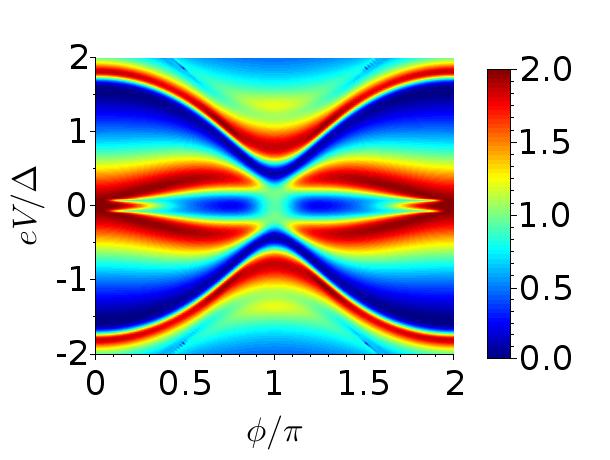}
 \includegraphics[width=4cm]{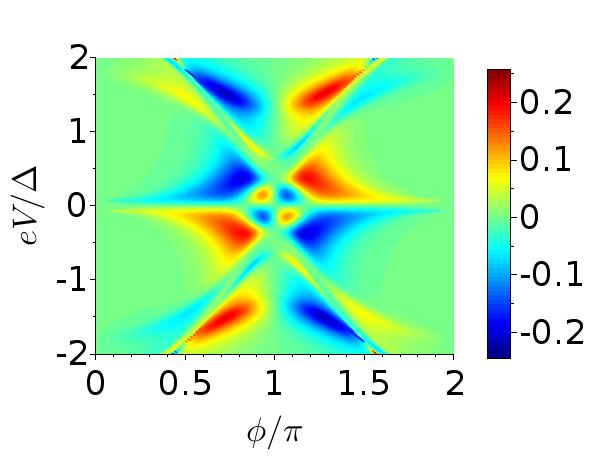}
 \caption{$G_{LL}$ (left panel) and $G_{RL}$ (right panel) in units of $e^2/h$ versus bias $eV$ and the phase difference $\phi=\phi_1-\phi_2$
 with $\phi_1=0$ for the choice of parameters: $t_0=10t$, $\De_0=0.99t_0$, $\mu_0=0$,
 $\De=0.1t$, $t'=3\De$, $\mu=0$, $t_{LM}=0.3t$, $t_{MK}=0.3t$, $t_{KR}=t$ and $L=40$.}\label{fig08}
\end{figure}

\begin{figure}
 \includegraphics[width=4cm]{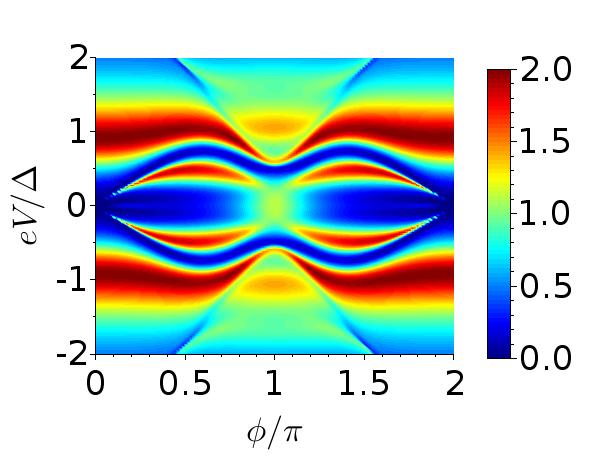}
 \includegraphics[width=4cm]{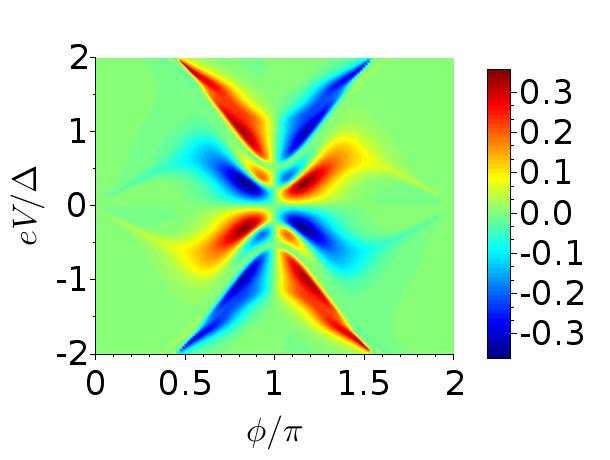}
 \caption{$G_{LL}$ (left panel) and $G_{RL}$ (right panel) in units of $e^2/h$ versus bias $eV$ and the phase difference $\phi=\phi_1-\phi_2$
 with $\phi_1=\pi$ for the choice of parameters: $t_0=10t$, $\De_0=0.99t_0$, $\mu_0=0$,
 $\De=0.1t$, $t'=3\De$, $\mu=0$, $t_{LM}=0.3t$, $t_{MK}=0.3t$, $t_{KR}=t$ and $L=40$.}\label{fig08-pi}
\end{figure}

\begin{figure}
 \includegraphics[width=4cm]{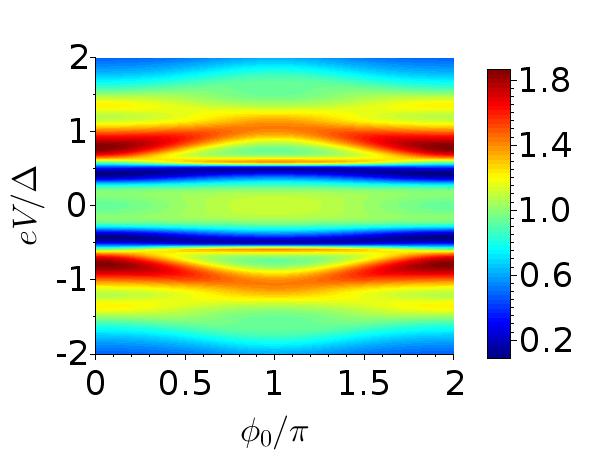}
 \includegraphics[width=4cm]{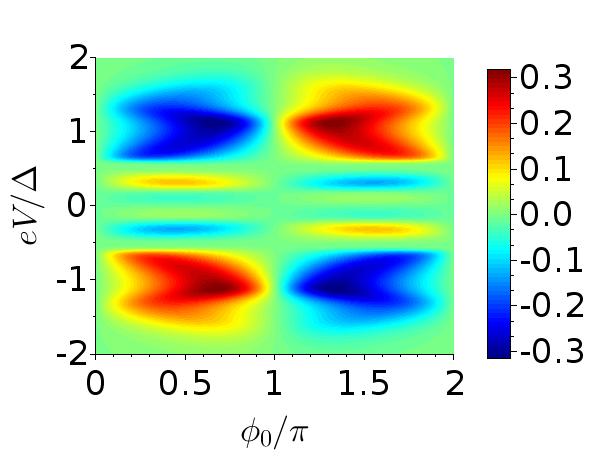}
 \caption{$G_{LL}$ (left panel) and $G_{RL}$ (right panel) in units of $e^2/h$ versus bias $eV$ and the overall phase $\phi_0$ where
 $\phi_1=\phi_0$ and $\phi_2=\pi+\phi_0$ for the choice of parameters: $t_0=10t$, $\De_0=0.99t_0$, $\mu_0=0$,
 $\De=0.1t$, $t'=3\De$, $\mu=0$, $t_{LM}=0.3t$, $t_{MK}=0.3t$, $t_{KR}=t$ and $L=40$.}\label{fig08-vsphi0}
\end{figure}

Now we turn to the dependence of the two conductances on the bias and the superconducting phase difference 
$\phi=(\phi_1-\phi_2)$, keeping $\phi_1$ constant. We fix $\mu=0$
and keep other parameters same as earlier. We see that the transconductance is enhanced in magnitude near $\phi=\pi$ in Fig.~\ref{fig08}.
The peak at zero bias in local conductance $G_{LL}$ for $\phi=0,2\pi$ is due to the MFs in the ladder. As $\phi$ increases from
$0$ to $\pi$, the MFs belonging to the two legs of the ladder hybridize and split in energy. This split in energy can be seen 
in the thick red lines originating at $(eV,\phi)=(0,0)$. As $\phi$ changes from $0$ to $\pi$, the bulk states of the two legs of the ladder
enter the gap $(-2\De,2\De)$ and form SAS. These are responsible for enhanced ET and enhanced CAR in the regions close to $\phi=\pi$.
Now, we change $\phi_1=\pi$ and $\phi_2=\pi-\phi$ and plot the dependence of the two conductances as functions of the bias $eV$ and the 
phase difference $\phi$ in Fig.~\ref{fig08-pi}. If we compare the local conductance plots of Fig.~\ref{fig08} and Fig.~\ref{fig08-pi}, 
we can see a clear contrast. Near zero $\phi$, there is a peak in $G_{LL}$ at $eV=\pm\De$ for $\phi_1=\pi$ while it is absent for 
$\phi_1=0$. Also, zero bias peak in
$G_{LL}$ at $\phi=0$ is absent for $\phi_1=\pi$ unlike the case $\phi_1=0$. This is an effect of interference between the MFs in the Kitaev
ladder and the MF in Kitaev chain. At $\phi=0$, even the Kitaev ladder hosts MFs. When $\phi_1=0$, the MF of the Kitaev ladder at zero energy is 
responsible for the zero bias peak. The peaks in local conductance at energies $\pm\De$ due to the fermion formed in the Kitaev chain are absent 
since the ladder allows for finite transmission through the evanescent modes. The peaks in local conductance at energies $\pm2\De$ are because the 
band bottom/top of the gapped SAS dispersion lies close to $\pm2\De$ and these modes cannot carry any current across due to zero group velocity.
But when $\phi_1=\pi$ and $\phi=0$, the Kitaev ladder and the Kitaev chain form exactly a $\pi$-Josephson junction and the subgap transport across a 
$\pi$-Josephson junction happens primarily by local Andreev reflection when the length of the Kitaev ladder is long~\cite{thaku15}.  
It is interesting to see that for both the cases, the transconductance has peaks and valleys near $\phi=\pi$. 
Motivated by the change in results when $\phi_1$ is changed, we study the dependence of the two conductances as a function of the overall 
phase keeping the phase difference the same. In Fig.~\ref{fig08-vsphi0} we plot the two conductances as functions of the bias and the overall 
phase of the ladder $\phi_0$ defined as $\phi_0=\phi_1=\phi_2-\pi$. Here, we maintain the phase difference between the two legs of the 
ladder to be $\pi$ since the ladder dispersion becomes gapless for this choice of $(\phi_1-\phi_2)$.
We see from this figure that the peaks in $G_{LL}$ at $eV\sim\pm\De$ have a slight variation as a function of $\phi_0$, while the
transconductance has thick regions of enhanced ET and enhanced CAR. 

\section{Discussion and Conclusion}
An interesting question that pops up in the analysis of the setup proposed is- `what
happens when the Kitaev chain is in the non-topological phase?' One way to take this limit is to set $t_0=10t$, 
$\De_0=0.99t_0$ and $\mu_0=3t_0$. Transport problem is solved in this limit taking other parameters same as that for Fig.~\ref{fig09}.
In this limit, the local conductance shows peaks for some values of the chemical potential $\mu$
and the bias $eV$, but the magnitude of the local conductance value is very small ($\sim10^{-6}$). The transconductance shows 
negative values for some range of chemical potential $\mu$ and the bias $eV$ but the magnitude is very small
($\sim10^{-7}$). This shows that electron reflection dominates for all values of the parameters of the Kitaev ladder. This is because 
the Kitaev chain acts like an insulator due to its discrete energy levels which are away from the band of the Kitaev ladder. Another way to
take the limit of non-topological phase for the Kitaev chain is to  set $\De_0=0$. In this limit, the Kitaev chain becomes a normal metal
when $\mu_0=0$ and $t_0=t$. The entire setup then becomes a Kitaev ladder connected to two normal metal leads with a quantum dot 
in between the left normal metal and the Kitaev ladder which has been essentially analyzed in Ref.~\cite{nehra19}. This shows negative
values of transconductance for a range of chemical potential and bias due to the SAS of the Kitaev ladder as expected. However in such
a non-topological phase of the Kitaev chain, there is no peak of magnitude $2e^2/h$ in local conductance when the
Kitaev chain is connected to a normal metal lead (this limit can be taken in our model by setting $t_{MK}=0$).  
The presence of MF is detected by peak in local conductance at the energy of the nonlocal fermion state formed (i.e., at 
$eV=\pm(t_0-\De_0)$ with a value $2e^2/h$)
when Kitaev chain is connected to a normal metal lead, 
while a negative value of transconductance (high in magnitude at the energies of the nonlocal fermion formed in the Kitaev chain) 
in the setup proposed here is the evidence of the nonlocality of MFs once the presence of MFs is detected. 

To conclude, we have seen that when a Kitaev ladder hosting SAS is connected to Kitaev chain hosting MFs, the transconductance shows large
positive and negative values for some choices of the parameters. Further, the transconductance shows periodic behavior which can be explained
by Fabry-P\'erot interference condition. Negative values of transconductance indicate enhanced crossed Andreev reflection.
We have thus shown that transconductance in the proposed setup can be used as a probe of the 
nonlocality of Majorana fermions.

\acknowledgements
The author thanks DST-INSPIRE Faculty Award (Faculty Reg. No.~:~IFA17-PH190)
for financial support. The author thanks Diptiman Sen and Subroto Mukerjee for comments on the manuscript. 

\bibliography{ref_mfcar}

\begin{thebibliography}{23}%
\makeatletter
\providecommand \@ifxundefined [1]{%
 \@ifx{#1\undefined}
}%
\providecommand \@ifnum [1]{%
 \ifnum #1\expandafter \@firstoftwo
 \else \expandafter \@secondoftwo
 \fi
}%
\providecommand \@ifx [1]{%
 \ifx #1\expandafter \@firstoftwo
 \else \expandafter \@secondoftwo
 \fi
}%
\providecommand \natexlab [1]{#1}%
\providecommand \enquote  [1]{``#1''}%
\providecommand \bibnamefont  [1]{#1}%
\providecommand \bibfnamefont [1]{#1}%
\providecommand \citenamefont [1]{#1}%
\providecommand \href@noop [0]{\@secondoftwo}%
\providecommand \href [0]{\begingroup \@sanitize@url \@href}%
\providecommand \@href[1]{\@@startlink{#1}\@@href}%
\providecommand \@@href[1]{\endgroup#1\@@endlink}%
\providecommand \@sanitize@url [0]{\catcode `\\12\catcode `\$12\catcode
  `\&12\catcode `\#12\catcode `\^12\catcode `\_12\catcode `\%12\relax}%
\providecommand \@@startlink[1]{}%
\providecommand \@@endlink[0]{}%
\providecommand \url  [0]{\begingroup\@sanitize@url \@url }%
\providecommand \@url [1]{\endgroup\@href {#1}{\urlprefix }}%
\providecommand \urlprefix  [0]{URL }%
\providecommand \Eprint [0]{\href }%
\providecommand \doibase [0]{http://dx.doi.org/}%
\providecommand \selectlanguage [0]{\@gobble}%
\providecommand \bibinfo  [0]{\@secondoftwo}%
\providecommand \bibfield  [0]{\@secondoftwo}%
\providecommand \translation [1]{[#1]}%
\providecommand \BibitemOpen [0]{}%
\providecommand \bibitemStop [0]{}%
\providecommand \bibitemNoStop [0]{.\EOS\space}%
\providecommand \EOS [0]{\spacefactor3000\relax}%
\providecommand \BibitemShut  [1]{\csname bibitem#1\endcsname}%
\let\auto@bib@innerbib\@empty
\bibitem [{\citenamefont {Kitaev}(2001)}]{kitaev2001unpaired}%
  \BibitemOpen
  \bibfield  {author} {\bibinfo {author} {\bibfnamefont {A.~Y.}\ \bibnamefont
  {Kitaev}},\ }\bibfield  {title} {\enquote {\bibinfo {title} {Unpaired
  majorana fermions in quantum wires},}\ }\href {\doibase
  10.1070/1063-7869/44/10s/s29} {\bibfield  {journal} {\bibinfo  {journal}
  {Phys.-Usp.}\ }\textbf {\bibinfo {volume} {44}},\ \bibinfo {pages} {131}
  (\bibinfo {year} {2001})}\BibitemShut {NoStop}%
\bibitem [{\citenamefont {Nayak}\ \emph {et~al.}(2008)\citenamefont {Nayak},
  \citenamefont {Simon}, \citenamefont {Stern}, \citenamefont {Freedman},\ and\
  \citenamefont {Das~Sarma}}]{nayak08}%
  \BibitemOpen
  \bibfield  {author} {\bibinfo {author} {\bibfnamefont {C.}~\bibnamefont
  {Nayak}}, \bibinfo {author} {\bibfnamefont {S.~H.}\ \bibnamefont {Simon}},
  \bibinfo {author} {\bibfnamefont {A.}~\bibnamefont {Stern}}, \bibinfo
  {author} {\bibfnamefont {M.}~\bibnamefont {Freedman}}, \ and\ \bibinfo
  {author} {\bibfnamefont {S.}~\bibnamefont {Das~Sarma}},\ }\bibfield  {title}
  {\enquote {\bibinfo {title} {Non-abelian anyons and topological quantum
  computation},}\ }\href {\doibase 10.1103/RevModPhys.80.1083} {\bibfield
  {journal} {\bibinfo  {journal} {Rev. Mod. Phys.}\ }\textbf {\bibinfo {volume}
  {80}},\ \bibinfo {pages} {1083--1159} (\bibinfo {year} {2008})}\BibitemShut
  {NoStop}%
\bibitem [{\citenamefont {Lutchyn}\ \emph {et~al.}(2010)\citenamefont
  {Lutchyn}, \citenamefont {Sau},\ and\ \citenamefont
  {Das~Sarma}}]{lutchyn2010majorana}%
  \BibitemOpen
  \bibfield  {author} {\bibinfo {author} {\bibfnamefont {R.~M.}\ \bibnamefont
  {Lutchyn}}, \bibinfo {author} {\bibfnamefont {J.~D.}\ \bibnamefont {Sau}}, \
  and\ \bibinfo {author} {\bibfnamefont {S.}~\bibnamefont {Das~Sarma}},\
  }\bibfield  {title} {\enquote {\bibinfo {title} {Majorana fermions and a
  topological phase transition in semiconductor-superconductor
  heterostructures},}\ }\href {\doibase 10.1103/PhysRevLett.105.077001}
  {\bibfield  {journal} {\bibinfo  {journal} {Phys. Rev. Lett.}\ }\textbf
  {\bibinfo {volume} {105}},\ \bibinfo {pages} {077001} (\bibinfo {year}
  {2010})}\BibitemShut {NoStop}%
\bibitem [{\citenamefont {Oreg}\ \emph {et~al.}(2010)\citenamefont {Oreg},
  \citenamefont {Refael},\ and\ \citenamefont {von Oppen}}]{oreg2010helical}%
  \BibitemOpen
  \bibfield  {author} {\bibinfo {author} {\bibfnamefont {Y.}~\bibnamefont
  {Oreg}}, \bibinfo {author} {\bibfnamefont {G.}~\bibnamefont {Refael}}, \ and\
  \bibinfo {author} {\bibfnamefont {F.}~\bibnamefont {von Oppen}},\ }\bibfield
  {title} {\enquote {\bibinfo {title} {Helical liquids and majorana bound
  states in quantum wires},}\ }\href {\doibase 10.1103/PhysRevLett.105.177002}
  {\bibfield  {journal} {\bibinfo  {journal} {Phys. Rev. Lett.}\ }\textbf
  {\bibinfo {volume} {105}},\ \bibinfo {pages} {177002} (\bibinfo {year}
  {2010})}\BibitemShut {NoStop}%
\bibitem [{\citenamefont {Mourik}\ \emph {et~al.}(2012)\citenamefont {Mourik},
  \citenamefont {Zuo}, \citenamefont {Frolov}, \citenamefont {Plissard},
  \citenamefont {Bakkers},\ and\ \citenamefont {Kouwenhoven}}]{mourik2012}%
  \BibitemOpen
  \bibfield  {author} {\bibinfo {author} {\bibfnamefont {V.}~\bibnamefont
  {Mourik}}, \bibinfo {author} {\bibfnamefont {K.}~\bibnamefont {Zuo}},
  \bibinfo {author} {\bibfnamefont {S.~M.}\ \bibnamefont {Frolov}}, \bibinfo
  {author} {\bibfnamefont {S.R.}\ \bibnamefont {Plissard}}, \bibinfo {author}
  {\bibfnamefont {E.~P. A.~M.}\ \bibnamefont {Bakkers}}, \ and\ \bibinfo
  {author} {\bibfnamefont {L.~P.}\ \bibnamefont {Kouwenhoven}},\ }\bibfield
  {title} {\enquote {\bibinfo {title} {Signatures of majorana fermions in
  hybrid superconductor-semiconductor nanowire devices},}\ }\href {\doibase
  10.1126/science.1222360} {\bibfield  {journal} {\bibinfo  {journal}
  {Science}\ }\textbf {\bibinfo {volume} {336}},\ \bibinfo {pages} {1003--1007}
  (\bibinfo {year} {2012})}\BibitemShut {NoStop}%
\bibitem [{\citenamefont {Das}\ \emph {et~al.}(2012)\citenamefont {Das},
  \citenamefont {Ronen}, \citenamefont {Most}, \citenamefont {Oreg},
  \citenamefont {Heiblum},\ and\ \citenamefont {Shtrikman}}]{Das2012}%
  \BibitemOpen
  \bibfield  {author} {\bibinfo {author} {\bibfnamefont {A.}~\bibnamefont
  {Das}}, \bibinfo {author} {\bibfnamefont {Y.}~\bibnamefont {Ronen}}, \bibinfo
  {author} {\bibfnamefont {Y.}~\bibnamefont {Most}}, \bibinfo {author}
  {\bibfnamefont {Y.}~\bibnamefont {Oreg}}, \bibinfo {author} {\bibfnamefont
  {M.}~\bibnamefont {Heiblum}}, \ and\ \bibinfo {author} {\bibfnamefont
  {H.}~\bibnamefont {Shtrikman}},\ }\bibfield  {title} {\enquote {\bibinfo
  {title} {Zero-bias peaks and splitting in an al-inas nanowire topological
  superconductor as a signature of majorana fermions},}\ }\href {\doibase
  10.1038/nphys2479} {\bibfield  {journal} {\bibinfo  {journal} {Nat. Phys.}\
  }\textbf {\bibinfo {volume} {8}},\ \bibinfo {pages} {887} (\bibinfo {year}
  {2012})}\BibitemShut {NoStop}%
\bibitem [{\citenamefont {Albrecht}\ \emph {et~al.}(2016)\citenamefont
  {Albrecht}, \citenamefont {Higginbotham}, \citenamefont {Madsen},
  \citenamefont {Kuemmeth}, \citenamefont {Jespersen}, \citenamefont
  {Nyg{\aa}rd}, \citenamefont {Krogstrup},\ and\ \citenamefont
  {Marcus}}]{albrecht2016}%
  \BibitemOpen
  \bibfield  {author} {\bibinfo {author} {\bibfnamefont {S.~M.}\ \bibnamefont
  {Albrecht}}, \bibinfo {author} {\bibfnamefont {A.~P.}\ \bibnamefont
  {Higginbotham}}, \bibinfo {author} {\bibfnamefont {M.}~\bibnamefont
  {Madsen}}, \bibinfo {author} {\bibfnamefont {F.}~\bibnamefont {Kuemmeth}},
  \bibinfo {author} {\bibfnamefont {T.~S.}\ \bibnamefont {Jespersen}}, \bibinfo
  {author} {\bibfnamefont {J.}~\bibnamefont {Nyg{\aa}rd}}, \bibinfo {author}
  {\bibfnamefont {P.}~\bibnamefont {Krogstrup}}, \ and\ \bibinfo {author}
  {\bibfnamefont {C.~M.}\ \bibnamefont {Marcus}},\ }\bibfield  {title}
  {\enquote {\bibinfo {title} {Exponential protection of zero modes in majorana
  islands},}\ }\href {\doibase 10.1038/nature17162} {\bibfield  {journal}
  {\bibinfo  {journal} {Nature}\ }\textbf {\bibinfo {volume} {531}},\ \bibinfo
  {pages} {206} (\bibinfo {year} {2016})}\BibitemShut {NoStop}%
\bibitem [{\citenamefont {Zhang}\ \emph {et~al.}(2018)\citenamefont {Zhang},
  \citenamefont {Liu}, \citenamefont {Gazibegovic}, \citenamefont {Xu},
  \citenamefont {Logan}, \citenamefont {Wang}, \citenamefont {van Loo},
  \citenamefont {Bommer}, \citenamefont {de~Moor}, \citenamefont {Car},
  \citenamefont {Op~het Veld}, \citenamefont {van Veldhoven}, \citenamefont
  {Koelling}, \citenamefont {Verheijen}, \citenamefont {Pendharkar},
  \citenamefont {Pennachio}, \citenamefont {Shojaei}, \citenamefont {Lee},
  \citenamefont {Palmstr{\o}m}, \citenamefont {Bakkers}, \citenamefont
  {Sarma},\ and\ \citenamefont {Kouwenhoven}}]{Zhang2018}%
  \BibitemOpen
  \bibfield  {author} {\bibinfo {author} {\bibfnamefont {H.}~\bibnamefont
  {Zhang}}, \bibinfo {author} {\bibfnamefont {C.-X.}\ \bibnamefont {Liu}},
  \bibinfo {author} {\bibfnamefont {S.}~\bibnamefont {Gazibegovic}}, \bibinfo
  {author} {\bibfnamefont {D.}~\bibnamefont {Xu}}, \bibinfo {author}
  {\bibfnamefont {J.~A.}\ \bibnamefont {Logan}}, \bibinfo {author}
  {\bibfnamefont {G.}~\bibnamefont {Wang}}, \bibinfo {author} {\bibfnamefont
  {N.}~\bibnamefont {van Loo}}, \bibinfo {author} {\bibfnamefont {J.~D.~S.}\
  \bibnamefont {Bommer}}, \bibinfo {author} {\bibfnamefont {M.~W.~A.}\
  \bibnamefont {de~Moor}}, \bibinfo {author} {\bibfnamefont {D.}~\bibnamefont
  {Car}}, \bibinfo {author} {\bibfnamefont {R.~L.~M.}\ \bibnamefont {Op~het
  Veld}}, \bibinfo {author} {\bibfnamefont {P.~J.}\ \bibnamefont {van
  Veldhoven}}, \bibinfo {author} {\bibfnamefont {S.}~\bibnamefont {Koelling}},
  \bibinfo {author} {\bibfnamefont {M.~A.}\ \bibnamefont {Verheijen}}, \bibinfo
  {author} {\bibfnamefont {M.}~\bibnamefont {Pendharkar}}, \bibinfo {author}
  {\bibfnamefont {D.~J.}\ \bibnamefont {Pennachio}}, \bibinfo {author}
  {\bibfnamefont {B.}~\bibnamefont {Shojaei}}, \bibinfo {author} {\bibfnamefont
  {J.~S.}\ \bibnamefont {Lee}}, \bibinfo {author} {\bibfnamefont {C.~J.}\
  \bibnamefont {Palmstr{\o}m}}, \bibinfo {author} {\bibfnamefont {E.~P. A.~M.}\
  \bibnamefont {Bakkers}}, \bibinfo {author} {\bibfnamefont {S.~D.}\
  \bibnamefont {Sarma}}, \ and\ \bibinfo {author} {\bibfnamefont {L.~P.}\
  \bibnamefont {Kouwenhoven}},\ }\bibfield  {title} {\enquote {\bibinfo {title}
  {Quantized majorana conductance},}\ }\href {\doibase 10.1038/nature26142}
  {\bibfield  {journal} {\bibinfo  {journal} {Nature}\ }\textbf {\bibinfo
  {volume} {556}},\ \bibinfo {pages} {74} (\bibinfo {year} {2018})}\BibitemShut
  {NoStop}%
\bibitem [{\citenamefont {Aguado}(2017)}]{aguado17}%
  \BibitemOpen
  \bibfield  {author} {\bibinfo {author} {\bibfnamefont {R.}~\bibnamefont
  {Aguado}},\ }\bibfield  {title} {\enquote {\bibinfo {title} {Majorana
  quasiparticles in condensed matter},}\ }\href@noop {} {\bibfield  {journal}
  {\bibinfo  {journal} {La Rivista del Nuovo Cimento}\ }\textbf {\bibinfo
  {volume} {40}},\ \bibinfo {pages} {523} (\bibinfo {year} {2017})}\BibitemShut
  {NoStop}%
\bibitem [{\citenamefont {Nilsson}\ \emph {et~al.}(2008)\citenamefont
  {Nilsson}, \citenamefont {Akhmerov},\ and\ \citenamefont
  {Beenakker}}]{nilsson2008}%
  \BibitemOpen
  \bibfield  {author} {\bibinfo {author} {\bibfnamefont {J.}~\bibnamefont
  {Nilsson}}, \bibinfo {author} {\bibfnamefont {A.R.}\ \bibnamefont
  {Akhmerov}}, \ and\ \bibinfo {author} {\bibfnamefont {C.~W.~J.}\ \bibnamefont
  {Beenakker}},\ }\bibfield  {title} {\enquote {\bibinfo {title} {Splitting of
  a cooper pair by a pair of majorana bound states},}\ }\href {\doibase
  10.1103/PhysRevLett.101.120403} {\bibfield  {journal} {\bibinfo  {journal}
  {Phys. Rev. Lett.}\ }\textbf {\bibinfo {volume} {101}},\ \bibinfo {pages}
  {120403} (\bibinfo {year} {2008})}\BibitemShut {NoStop}%
\bibitem [{\citenamefont {L\"u}\ \emph {et~al.}(2012)\citenamefont {L\"u},
  \citenamefont {Lu},\ and\ \citenamefont {Shen}}]{Lu12}%
  \BibitemOpen
  \bibfield  {author} {\bibinfo {author} {\bibfnamefont {H.-F.}\ \bibnamefont
  {L\"u}}, \bibinfo {author} {\bibfnamefont {H.-Z.}\ \bibnamefont {Lu}}, \ and\
  \bibinfo {author} {\bibfnamefont {S.-Q.}\ \bibnamefont {Shen}},\ }\bibfield
  {title} {\enquote {\bibinfo {title} {Nonlocal noise cross correlation
  mediated by entangled majorana fermions},}\ }\href {\doibase
  10.1103/PhysRevB.86.075318} {\bibfield  {journal} {\bibinfo  {journal} {Phys.
  Rev. B}\ }\textbf {\bibinfo {volume} {86}},\ \bibinfo {pages} {075318}
  (\bibinfo {year} {2012})}\BibitemShut {NoStop}%
\bibitem [{\citenamefont {Liu}\ \emph {et~al.}(2013)\citenamefont {Liu},
  \citenamefont {Zhang},\ and\ \citenamefont {Law}}]{Liu13}%
  \BibitemOpen
  \bibfield  {author} {\bibinfo {author} {\bibfnamefont {J.}~\bibnamefont
  {Liu}}, \bibinfo {author} {\bibfnamefont {F.-C.}\ \bibnamefont {Zhang}}, \
  and\ \bibinfo {author} {\bibfnamefont {K.~T.}\ \bibnamefont {Law}},\
  }\bibfield  {title} {\enquote {\bibinfo {title} {Majorana fermion induced
  nonlocal current correlations in spin-orbit coupled superconducting wires},}\
  }\href {\doibase 10.1103/PhysRevB.88.064509} {\bibfield  {journal} {\bibinfo
  {journal} {Phys. Rev. B}\ }\textbf {\bibinfo {volume} {88}},\ \bibinfo
  {pages} {064509} (\bibinfo {year} {2013})}\BibitemShut {NoStop}%
\bibitem [{\citenamefont {Zocher}\ and\ \citenamefont
  {Rosenow}(2013)}]{Zocher13}%
  \BibitemOpen
  \bibfield  {author} {\bibinfo {author} {\bibfnamefont {B.}~\bibnamefont
  {Zocher}}\ and\ \bibinfo {author} {\bibfnamefont {B.}~\bibnamefont
  {Rosenow}},\ }\bibfield  {title} {\enquote {\bibinfo {title} {Modulation of
  majorana-induced current cross-correlations by quantum dots},}\ }\href
  {\doibase 10.1103/PhysRevLett.111.036802} {\bibfield  {journal} {\bibinfo
  {journal} {Phys. Rev. Lett.}\ }\textbf {\bibinfo {volume} {111}},\ \bibinfo
  {pages} {036802} (\bibinfo {year} {2013})}\BibitemShut {NoStop}%
\bibitem [{\citenamefont {Prada}\ \emph {et~al.}(2017)\citenamefont {Prada},
  \citenamefont {Aguado},\ and\ \citenamefont {San-Jose}}]{Prada17}%
  \BibitemOpen
  \bibfield  {author} {\bibinfo {author} {\bibfnamefont {E.}~\bibnamefont
  {Prada}}, \bibinfo {author} {\bibfnamefont {R.}~\bibnamefont {Aguado}}, \
  and\ \bibinfo {author} {\bibfnamefont {P.}~\bibnamefont {San-Jose}},\
  }\bibfield  {title} {\enquote {\bibinfo {title} {Measuring majorana
  nonlocality and spin structure with a quantum dot},}\ }\href {\doibase
  10.1103/PhysRevB.96.085418} {\bibfield  {journal} {\bibinfo  {journal} {Phys.
  Rev. B}\ }\textbf {\bibinfo {volume} {96}},\ \bibinfo {pages} {085418}
  (\bibinfo {year} {2017})}\BibitemShut {NoStop}%
\bibitem [{\citenamefont {Liu}\ \emph {et~al.}(2017)\citenamefont {Liu},
  \citenamefont {Song}, \citenamefont {Sun},\ and\ \citenamefont
  {Xie}}]{liu17}%
  \BibitemOpen
  \bibfield  {author} {\bibinfo {author} {\bibfnamefont {J}~\bibnamefont
  {Liu}}, \bibinfo {author} {\bibfnamefont {J.}~\bibnamefont {Song}}, \bibinfo
  {author} {\bibfnamefont {Q.-F.}\ \bibnamefont {Sun}}, \ and\ \bibinfo
  {author} {\bibfnamefont {X.~C.}\ \bibnamefont {Xie}},\ }\bibfield  {title}
  {\enquote {\bibinfo {title} {Even-odd interference effect in a topological
  superconducting wire},}\ }\href {\doibase 10.1103/PhysRevB.96.195307}
  {\bibfield  {journal} {\bibinfo  {journal} {Phys. Rev. B}\ }\textbf {\bibinfo
  {volume} {96}},\ \bibinfo {pages} {195307} (\bibinfo {year}
  {2017})}\BibitemShut {NoStop}%
\bibitem [{\citenamefont {Nehra}\ \emph {et~al.}(2019)\citenamefont {Nehra},
  \citenamefont {Bhakuni}, \citenamefont {Sharma},\ and\ \citenamefont
  {Soori}}]{nehra19}%
  \BibitemOpen
  \bibfield  {author} {\bibinfo {author} {\bibfnamefont {R.}~\bibnamefont
  {Nehra}}, \bibinfo {author} {\bibfnamefont {D.~S.}\ \bibnamefont {Bhakuni}},
  \bibinfo {author} {\bibfnamefont {A.}~\bibnamefont {Sharma}}, \ and\ \bibinfo
  {author} {\bibfnamefont {A.}~\bibnamefont {Soori}},\ }\bibfield  {title}
  {\enquote {\bibinfo {title} {Enhancement of crossed andreev reflection in a
  kitaev ladder connected to normal metal leads},}\ }\href {\doibase
  10.1088/1361-648x/ab2403} {\bibfield  {journal} {\bibinfo  {journal} {Jour.
  Phys.: Cond. Matt.}\ }\textbf {\bibinfo {volume} {31}},\ \bibinfo {pages}
  {345304} (\bibinfo {year} {2019})}\BibitemShut {NoStop}%
\bibitem [{\citenamefont {Deutscher}\ and\ \citenamefont
  {Feinberg}(2000)}]{deutscher2000}%
  \BibitemOpen
  \bibfield  {author} {\bibinfo {author} {\bibfnamefont {G.}~\bibnamefont
  {Deutscher}}\ and\ \bibinfo {author} {\bibfnamefont {D.}~\bibnamefont
  {Feinberg}},\ }\bibfield  {title} {\enquote {\bibinfo {title} {Coupling
  superconducting-ferromagnetic point contacts by andreev reflections},}\
  }\href {\doibase 10.1063/1.125796} {\bibfield  {journal} {\bibinfo  {journal}
  {Appl. Phys. Lett.}\ }\textbf {\bibinfo {volume} {76}},\ \bibinfo {pages}
  {487--489} (\bibinfo {year} {2000})}\BibitemShut {NoStop}%
\bibitem [{\citenamefont {He}\ \emph {et~al.}(2014)\citenamefont {He},
  \citenamefont {Wu}, \citenamefont {Choy}, \citenamefont {Liu}, \citenamefont
  {Tanaka},\ and\ \citenamefont {Law}}]{he14}%
  \BibitemOpen
  \bibfield  {author} {\bibinfo {author} {\bibfnamefont {J.~J.}\ \bibnamefont
  {He}}, \bibinfo {author} {\bibfnamefont {J.}~\bibnamefont {Wu}}, \bibinfo
  {author} {\bibfnamefont {T.~P.}\ \bibnamefont {Choy}}, \bibinfo {author}
  {\bibfnamefont {X.~J.}\ \bibnamefont {Liu}}, \bibinfo {author} {\bibfnamefont
  {Y.}~\bibnamefont {Tanaka}}, \ and\ \bibinfo {author} {\bibfnamefont {K.~T.}\
  \bibnamefont {Law}},\ }\bibfield  {title} {\enquote {\bibinfo {title}
  {Correlated spin currents generated by resonant-crossed andreev reflections
  in topological superconductors},}\ }\href {\doibase 10.1038/ncomms4232}
  {\bibfield  {journal} {\bibinfo  {journal} {Nat. Commun.}\ }\textbf {\bibinfo
  {volume} {5}},\ \bibinfo {pages} {3232} (\bibinfo {year} {2014})}\BibitemShut
  {NoStop}%
\bibitem [{\citenamefont {Levy~Yeyati}\ \emph {et~al.}(2007)\citenamefont
  {Levy~Yeyati}, \citenamefont {Bergeret}, \citenamefont {Martin-Rodero},\ and\
  \citenamefont {Klapwijk}}]{yeyati07}%
  \BibitemOpen
  \bibfield  {author} {\bibinfo {author} {\bibfnamefont {A.}~\bibnamefont
  {Levy~Yeyati}}, \bibinfo {author} {\bibfnamefont {F.~S.}\ \bibnamefont
  {Bergeret}}, \bibinfo {author} {\bibfnamefont {A.}~\bibnamefont
  {Martin-Rodero}}, \ and\ \bibinfo {author} {\bibfnamefont {T.~M.}\
  \bibnamefont {Klapwijk}},\ }\bibfield  {title} {\enquote {\bibinfo {title}
  {Entangled andreev pairs and collective excitations in nanoscale
  superconductors},}\ }\href {\doibase 10.1038/nphys621} {\bibfield  {journal}
  {\bibinfo  {journal} {Nat. Phys.}\ }\textbf {\bibinfo {volume} {3}},\
  \bibinfo {pages} {455} (\bibinfo {year} {2007})}\BibitemShut {NoStop}%
\bibitem [{\citenamefont {Soori}\ and\ \citenamefont
  {Mukerjee}(2017)}]{soori17}%
  \BibitemOpen
  \bibfield  {author} {\bibinfo {author} {\bibfnamefont {A.}~\bibnamefont
  {Soori}}\ and\ \bibinfo {author} {\bibfnamefont {S.}~\bibnamefont
  {Mukerjee}},\ }\bibfield  {title} {\enquote {\bibinfo {title} {Enhancement of
  crossed andreev reflection in a superconducting ladder connected to normal
  metal leads},}\ }\href {\doibase 10.1103/PhysRevB.95.104517} {\bibfield
  {journal} {\bibinfo  {journal} {Phys. Rev. B}\ }\textbf {\bibinfo {volume}
  {95}},\ \bibinfo {pages} {104517} (\bibinfo {year} {2017})}\BibitemShut
  {NoStop}%
\bibitem [{\citenamefont {Thakurathi}\ \emph {et~al.}(2015)\citenamefont
  {Thakurathi}, \citenamefont {Deb},\ and\ \citenamefont {Sen}}]{thaku15}%
  \BibitemOpen
  \bibfield  {author} {\bibinfo {author} {\bibfnamefont {M.}~\bibnamefont
  {Thakurathi}}, \bibinfo {author} {\bibfnamefont {O.}~\bibnamefont {Deb}}, \
  and\ \bibinfo {author} {\bibfnamefont {D.}~\bibnamefont {Sen}},\ }\bibfield
  {title} {\enquote {\bibinfo {title} {Majorana modes and transport across
  junctions of superconductors and normal metals},}\ }\href {\doibase
  10.1088/0953-8984/27/27/275702} {\bibfield  {journal} {\bibinfo  {journal}
  {Jour. Phys.: Cond. Matt.}\ }\textbf {\bibinfo {volume} {27}},\ \bibinfo
  {pages} {275702} (\bibinfo {year} {2015})}\BibitemShut {NoStop}%
\bibitem [{\citenamefont {Datta}(1995)}]{datta1995}%
  \BibitemOpen
  \bibfield  {author} {\bibinfo {author} {\bibfnamefont {S.}~\bibnamefont
  {Datta}},\ }\href@noop {} {\emph {\bibinfo {title} {Electronic transport in
  mesoscopic systems}}}\ (\bibinfo  {publisher} {Cambridge University Press,
  Cambridge},\ \bibinfo {year} {1995})\BibitemShut {NoStop}%
\bibitem [{\citenamefont {Soori}\ \emph {et~al.}(2012)\citenamefont {Soori},
  \citenamefont {Das},\ and\ \citenamefont {Rao}}]{soori12}%
  \BibitemOpen
  \bibfield  {author} {\bibinfo {author} {\bibfnamefont {A.}~\bibnamefont
  {Soori}}, \bibinfo {author} {\bibfnamefont {S.}~\bibnamefont {Das}}, \ and\
  \bibinfo {author} {\bibfnamefont {S.}~\bibnamefont {Rao}},\ }\bibfield
  {title} {\enquote {\bibinfo {title} {Magnetic-field-induced fabry-p\'erot
  resonances in helical edge states},}\ }\href {\doibase
  10.1103/PhysRevB.86.125312} {\bibfield  {journal} {\bibinfo  {journal} {Phys.
  Rev. B}\ }\textbf {\bibinfo {volume} {86}},\ \bibinfo {pages} {125312}
  (\bibinfo {year} {2012})}\BibitemShut {NoStop}%
\end{thebibliography}%

\end{document}